\documentclass[10pt,journal,letterpaper,compsoc]{IEEEtran}
\ifCLASSOPTIONcompsoc
\else
\fi
\ifCLASSINFOpdf
\else
\fi
\usepackage{cite}
\usepackage{graphicx}
\usepackage{psfrag}
\usepackage{subfigure}
\usepackage{url}
\usepackage{amsmath}
\usepackage[normalem]{ulem}
\usepackage{epsfig,colortbl}
\usepackage{amssymb,comment}
\usepackage{enumerate}
\usepackage{times}
\usepackage{multirow,multicol}
\usepackage[ruled, vlined]{algorithm2e}
\newcounter{ctr}

\newtheorem{theorem}{Theorem}

\makeatletter

\newcommand{\be}{\begin{eqnarray}}
\newcommand{\ee}{\end{eqnarray}}
\newcommand{\nn}{\nonumber}

\newcommand{\bm}{\boldmath}

\newcommand{\cc}{\mbox{\bm $c$}}

\newcommand{\PP}{\mbox{\bm $P$}}
\newcommand{\uu}{\mbox{\bm $u$}}
\newcommand{\ccc}{{\bf c}}

\newcommand{\rr}{\mbox{\bm $r$}}

\newcommand{\G}{\mbox{\bm $G$}}

\renewcommand\paragraph{\@startsection{paragraph}{4}{\z@}%
    {1.5ex plus .2ex minus .3ex}%
            {-0em}%
                        {\normalsize\bf}}

\begin{document}
%
\title{Progressive Decoding for Data Availability and Reliability in Distributed Networked Storage}
%
%
%
%

\author{Yunghsiang~S.~Han,~\IEEEmembership{Senior Member,~IEEE,}
	Soji~Omiwade,~\IEEEmembership{Member,~IEEE,}
        and~Rong~Zheng,~\IEEEmembership{Member,~IEEE}
\IEEEcompsocitemizethanks{\IEEEcompsocthanksitem Han is with the Department of Electrical  Engineering, National Taiwan University of Science and Technology, Taiwan, R.O.C. (e-mail: yshan@mail.ntust.edu.tw)
\IEEEcompsocthanksitem Omiwade and Zheng are with the Department of Computer Science, University of Houston, Houston, TX 77204 USA (e-mail: 
\{ooo00a,rzheng\}@cs.uh.edu.}
}

\IEEEcompsoctitleabstractindextext{%
\begin{abstract}
To harness the ever growing capacity and decreasing cost of storage,
providing an abstraction of dependable storage in the presence of
crash-stop and Byzantine failures is compulsory.  We propose a
decentralized Reed Solomon coding mechanism with minimum communication
overhead.  Using a progressive data retrieval scheme, a data collector
contacts only the necessary number of storage nodes needed to guarantee
data integrity.  The scheme gracefully adapts the cost of successful data
retrieval to the number of storage node failures. Moreover, by
leveraging the Welch-Berlekamp algorithm, it avoids unnecessary
computations. Compared to the state-of-the-art decoding scheme, the
implementation and evaluation results show that our progressive data
retrieval scheme has up to $35$ times better computation performance
for low Byzantine node rates.  Additionally, the communication cost in
data retrieval is derived analytically and corroborated by Monte-Carlo
simulation results. Our implementation is flexible in that the level of
redundancy it provides is independent of the number of data generating
nodes, a requirement for distributed storage systems.
\end{abstract}


\begin{keywords}
Reliability, Availability, Fault tolerance, Error control codes
\end{keywords}}

\maketitle

\IEEEdisplaynotcompsoctitleabstractindextext

%
\IEEEpeerreviewmaketitle

\section{Introduction}
\label{sect:intro}

\IEEEPARstart{C}{ost} of storage for data availability over networks has decreased drastically over the years.  Companies such
as Google and Amazon offer TB of online storage for free or at a very low
cost.  Also, low-power storage media are widely used in embedded devices
or mobile computers. However, to harness the ever growing capacity and
decreasing cost of distributed storage for persistent data availability, a number of challenges need to be
addressed, (i) volatility of storage due to network
disconnectivity, varying administrative restriction or user preferences, and
nodal mobility (of mobile devices); (ii) (partial) failures of storage devices.
For example, flash media are known to be engineered to trade-off error
probabilities for cost reduction; (iii) software bugs or malicious attacks, where an
adversary manages to compromise enough storage nodes to guarantee that integrity cannot be guaranteed.

To ensure availability and data integrity despite failure or compromise of storage nodes,
survivable storage systems spread data redundantly across a set of distributed
storage nodes. At the core of a survivable storage system is a coding scheme
that maps information bits to stored bits, and vice versa. 
The unit of such mapping are referred to as \emph{symbols} in this paper. A $(n,k)$ coding is
defined by the following two primitives:
\begin{itemize}
\item[-] {\bf encode} $\mathbf{c} = (\mathbf{u}, n, k)$ takes as input 
$k$ information symbols $\mathbf{u} = [u_0, u_1, \ldots, u_{k-1}]$ and
returns a coded
vector $\mathbf{c} = [c_0, c_1, \ldots, c_{n-1}]$.
The coded symbols are stored on storage nodes, one per node.
\item[-] {\bf decode} $\mathbf{u} = (\mathbf{r}, n, k)$ accesses a
subset of storage nodes and returns the original $k$ information symbols from possibly
corrupted symbols.
\end{itemize}

Most existing approaches to survivable and dependable storage assume crash-stop behaviors.
That is, a storage device becomes unavailable if failed (also called ``erasure").
Solutions such as various RAID configurations~\cite{raid} and their extensions
are engineered for high read and write data throughput. In this case, typically
low-complexity (replication or XOR-based) coding mechanisms are employed to
recover the original data from limited degree of erasure. We argue that Byzantine failures, where
devices fail in arbitrary manner and cannot be \emph{trusted}, are becoming more
pertinent with the prevalence of cheap storage devices, software bugs and
malicious attacks. Efficient encode and decode primitives that can detect data
corruption and handle Byzantine failures serve as a fundamental building block to
support higher level abstractions such as multi-reader multi-writer atomic
register~\cite{Goodson04} and digital fingerprints~\cite{Krawczyk93} in dependable distributed systems. 

For fixed error correction capability, the efficiency of encode and decode
primitives can be evaluated by three metrics, i) {\it storage overhead}
measured as the ratio between the number of storage symbols and total
information symbols ($n/k$); ii) {\it encoding and decoding computation time};
and iii) {\it communication overhead} measured in the number of bits
transferred in the network for encode and decode. Communication overhead is of
much importance in wide-area and/or low-bandwidth storage systems. Even though Reed-Solomon (RS) codes have been used for distributed storage for a single system, they have been found unsuitable for distributed networked storage due to their centralized nature~\cite{DIM06} and high communication overhead~\cite{lin2007dpl}. However, by encoding data at each data node, we found that RS codes can avoid the above disadvantages and provide better performance in almost every aspect than existing storage schemes. Hence, in this
paper, we propose a novel solution to spreading redundant information efficiently
across distributed storage nodes using incremental RS decoding.
By virtue of RS codes, our scheme is storage optimal. The key novelty of the proposed
approach lies in a progressive data retrieval procedure, which retrieves
just enough data from live storage nodes, and performs decoding
incrementally.  As a result, both communication and computation cost are
minimized, and adapt to the degree of errors in the system. We
provide a theoretical characterization of the communication cost and success
rate of data retrieval using the proposed scheme in presence of arbitrary
errors in the system. Our implementation studies demonstrate up to 20 times
speed-up of the progressive data retrieval scheme in computation time, relative
to a classical scheme. Moreover, the proposed scheme is comparable to that of a
genie-aid decoding process, which assumes knowledge of failure modes of storage
nodes.

In this paper, we make the following contributions:
\begin{itemize}
\item Design of a novel progressive data retrieval algorithm that is storage
and communication optimal, and computationally efficient. It handles Byzantine
failures in storage nodes gracefully as the probability of failures increases.
\item Development of an analytical model to evaluate the communication cost of
our data retrieval algorithm.
\item Eliminate the need for the number of data nodes, $k$, to equal the number of
information symbols, $\hat{k}$--a constraint that is restrictive and unrealistic
for general distributed storage systems.
\end{itemize}

The rest of the paper is organized as follows. Related work is given in
Section~\ref{sect:related}. The progressive data retrieval scheme is presented
in Section~\ref{sect:progressive}, with the details of the incremental RS
decoding algorithm in Section~\ref{sect:algo}.  An analysis of our coding,
communication and success rate complexity is provided in
Section~\ref{sect:complexity}, and Section~\ref{sect:era_comp} compares our scheme with leading erasure coding protocols. Evaluation results are presented in
Section~\ref{sect:eval}.  Finally, we
conclude the paper in Section~\ref{sect:conc}.
\section{Background and Related Work}
\label{sect:related}
\begin{figure*}[thp]
\begin{center}
\includegraphics[width=6in]{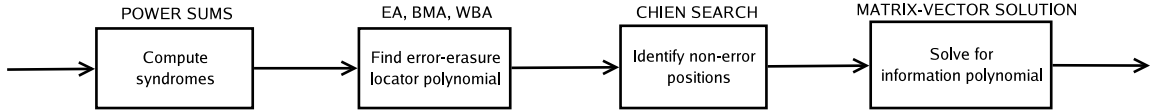}
\caption{Block diagram of RS decoding. Above each block, the corresponding existing algorithms are indicated.}
\label{fig:rs_decode}
\end{center}
\end{figure*}
In storage systems, ensuring reliability and data integrity requires the introduction of
redundancy. A file is divided into $k$ symbols, encoded into $n$ coded  symbols and
stored at $n$ nodes. One important metric of coding efficiency is the
redundancy-reliability trade off defined as $n/k$. The simplest form of redundancy
is replication.  As a generalization of replication, erasure coding offers
better storage efficiency. The Maximum Distance Separable (MDS) codes are
optimal as it provides largest separation among code words, and an $(n, k)-$MDS
code will be able to recover from any $v$ errors if 
$v \le \lfloor{\frac{n-k-s}{2}}\rfloor$,
where $s$ is the number of erasures (or irretrievable symbols).
\subsection{Reed-Solomon codes}
RS codes are the most well-known class of MDS codes. They not only can recover data when nodes fail, but can guarantee
recovery when a certain subset of nodes are Byzantine. RS codes operate on symbols of
$m$ bits. An $(n,k)$ RS code is a linear code, with each symbol in
$GF(2^{m})$, and parameters $n=2^{m}-1$ and $n-k=2t~,$ where $n$ is the total
number of symbols in a codeword, $k$ is the total number of information
symbols, and $t$ is the symbol-error-correcting capability of the code.
\paragraph*{Encoding}
Let the sequence of $k$ information symbols in $GF(2^{m})$ be
$\uu=[u_0,u_1,\ldots,u_{k-1}]$ and $u(x)$ be the information polynomial of $\uu$
represented as
$$u(x)=u_0+u_1x+\cdots+u_{k-1}x^{k-1}~.$$
The codeword polynomial, $c(x)$, corresponding to $u(x)$ can be
encoded as
$$c(x)=u(x)g(x)~,$$
where $g(x)$ is a generator polynomial of the RS code. It is
well-known that $g(x)$ can be obtained as
\begin{eqnarray}
\label{g(x)}
g(x)&=&(x-\alpha^b)(x-\alpha^{b+1})\cdots(x-\alpha^{b+2t-1})\nonumber\\
&=&g_0+g_1x+g_2x^2+\cdots+g_{2t}x^{2t}~,
\end{eqnarray}
where $\alpha$ is a primitive element in $GF(2^{m})$, $b$ an arbitrary integer, and $g_i\in
GF(2^{m})$.
\paragraph*{Decoding}
The decoding process of RS codes is more complex. Complete description of decoding of RS codes can be found
in~\cite{MOO05}.

Let $r(x)$ be the received polynomial and
$r(x)=c(x)+e(x)+\gamma(x)=c(x)+\lambda(x)$, where $e(x)=
\sum_{j=0}^{n-1}e_jx^j$ is the error polynomial, $\gamma(x)=
\sum_{j=0}^{n-1}\gamma_jx^j$  the erasure polynomial, and
$\lambda(x)=\sum_{j=0}^{n-1}\lambda_jx^j=e(x)+\gamma(x)$  the
errata polynomial. Note that $g(x)$ and (hence) $c(x)$ have
$\alpha^b,\alpha^{b+1},\ldots,\alpha^{b+2t-1}$ as roots. This property is used
to determine the error locations and recover the information symbols.

The basic procedure of RS decoding is shown in Figure~\ref{fig:rs_decode}. The
last step of the decoding procedure involves solving a linear set of equations,
and can be made efficient by the use of Vandermonde generator matrices~\cite{william1988numerical}.

In $GF(2^m)$, addition is equivalent to bit-wise exclusive-or (XOR), and
multiplication is typically implemented with multiplication tables or discrete
logarithm tables.  To reduce the complexity of multiplication, Cauchy
Reed-Solomon (CRS) codes~\cite{Blomer95anxor-based} have been proposed to use a different
construction of the generator matrix, and convert multiplications to XOR
operations for erasure.  However, CRS codes incur the same complexity as RS codes for
error corrections. 
\subsection{Existing work}
Several XOR-based erasure codes (in a field of
GF(2))~\cite{corbett4rdp,blaum1999ldm,blaum:eca,lin2007dpl} have been used in
storage systems.  In RAID-6 systems, each disk is partitioned into strips of
fixed size. Two parity strips are computed using one strip from each data disk,
forming a stripe together with the data strips.  EVEN-ODD\cite{blaum:eca}, Row
Diagonal Parity (RDP)\cite{corbett4rdp}, and Minimal Density RAID-6
codes\cite{blaum1999ldm}  use XOR operations, and are specific to RAID-6.
A detailed comparison of the encoding and decoding
performance of several open-source erasure coding libraries for storage is provided\cite{Plank09}.
We mention that the gain in computation efficiency of XOR-based erasure codes is achieved by
trading off fault tolerance. 
Our progressive data retrieval algorithm--however--
can tolerate as much fault--according to the configuration of the code's robustness--as is needed 
\emph{and} is highly efficient both in computation and communication costs.
Moreover, RAID-6 systems can recover from the loss of
exactly two disks but cannot handle Byzantine failures, thereby eliminating the application of such
systems for sensor networks. 

In the context of network storage for wireless sensor networks,
randomized linear codes~\cite{DIM06} and fountain codes~\cite{lin2007dpl} have
been applied with the objective that a data collector can retrieve unit data
from each of $k$ data sources by accessing any $k$ out $n$ storage nodes,
and thus up to $n-k$ crash-stop node failures can be tolerated.  However, such schemes
cannot recover from data modifications in the field. Compared to erasure
based solutions, the key distinctions are i) coding is done at the storage
nodes rather than at the data source, and ii) each storage node only has unit
capacity. Later, we provide a reference implementation of a single data
collector problem using the proposed primitives. Our evaluation studies shows that
our implementation outperforms the distributed storage scheme based on random
linear network coding in almost all metrics. 
\section{Progressive Data Retrieval}
\label{sect:progressive}
We use the abstractions of a data node which is a source of information that
must be stored, and a storage node which corresponds to a storage device.
Nodes are subject to both crash-stop failures, where data cannot be accessed
and Byzantine failures, where arbitrary data may be returned.  The
communication cost of transferring one unit of data from the data source to a
storage node is assumed to be constant independent of the location of the
storage node.
Unlike existing decentralized schemes for distributed networked storage, the message length in each encoding process of the proposed scheme is not tied with the number of data  node, $k$. Hence, the RS code used in the proposed scheme is denoted as an $(n,\hat{k})$ code. The scheme given in~\cite{Han10-Infocom} is a special case when $\hat{k}=k$. It will be shown that the value of $\hat{k}$ affects the storage efficiency and the communication cost.
\subsection{Data storage}
\label{sect:data_storage}
The data storage scheme consists of two steps. First, for data integrity, a
message authentication code (MAC) is added to each data block generated by a data node
before it is encoded. One-way hash functions such as MD5, SHA-1, SHA-2 can
be used. For simplicity, we adopt  CRC code for error detection with $r$
redundant bits~\cite{MOO05,REE99}. The portion of
errors that cannot be detected by a CRC code is dependent only on its number of
redundant bits. That is, a CRC code with $r$ redundant bits {\it cannot} detect
$(\frac{1}{2^r})100\%$ portion of errors.  If $T_0$ is the size of the original data with header information, then the size of the resulting data with CRC is $T = T_0 + r$. It is easy to see that the CRC overhead can be amortized by using large data blocks.  

In the second step, a data block is partitioned into information symbols of
length $m$ bits and RS codes are applied. The data node divides its
data into $\lceil T/m\rceil$ symbols such
that each symbol represents an element in $GF(2^m)$. Next 
the $\lceil T/m\rceil$ symbols are divided into $\left\lceil\frac{\lceil
T/m\rceil}{\hat{k}}\right\rceil$ information groups each of $\hat{k}$
symbols.\footnote{CRC codes are added to every information group. Hence, the data size $T$ includes those bits added by CRC codes. The last information group may have less than $\hat{k}$ symbols.
In this case, zeros are appended during the encoding procedure.} Let
$\hat{k}$ symbols of the $i$-th group be the
components in information vector
$\uu_i=[u_{i0},u_{i1},\ldots,u_{i(\hat{k}-1)}]$,
where $1\le i\le \left\lceil\frac{\lceil
T/m\rceil}{\hat{k}}\right\rceil$.   The node encodes $\uu_i$ into
$\cc_i=[c_{i0},c_{i1},\ldots,c_{i(n-1)}]$
with $n$ symbols as $$\cc_i=\uu_i\G,$$
where 
\begin{eqnarray}
\G=\left[\begin{array}{ccccc} 1&1&1&\cdots&1\\
\alpha&\alpha^2&\alpha^3&\cdots&\alpha^n\\
\alpha^2&(\alpha^2)^2&(\alpha^3)^2&\cdots&(\alpha^n)^2\\
&&&\vdots&\\
\alpha^{\hat{k}-1}&(\alpha^2)^{\hat{k}-1}&(\alpha^3)^{\hat{k}-1}&\cdots&(\alpha^n)^{\hat{k}-1}
\end{array}\right]~.
\label{eq:generator}
\end{eqnarray}
Recall that $\alpha$ is a primitive element (generator) of $GF(2^m)$ which can be
determined in advance. The data node then packs all $c_{i,j}$, $0\le
i\le \left\lceil\frac{\lceil T/m\rceil}{\hat{k}}\right\rceil$, and sends them with
their index $j$ to storage node $(j+1)$ via the network. 
\subsection{Data retrieval}
To reconstruct the source data, a collector needs to access sufficient number
of storage nodes to ensure data integrity. Among $n$ storage nodes, let the
number of erasures, which includes the number of crash-stop nodes and the
number of nodes that have not been accessed, be $s$. Identity of crash-stop
nodes can be determined by the use of keep-alive messages.  Additionally, there
are $v$ nodes with Byzantine failures. Neither $v$ nor the identity of these
nodes are known to the data collector. 

$\G$ given in~(\ref{eq:generator}) is a generator
matrix of a RS code~\cite{MOO05} and thus an error-erasure decoding algorithm
can recover all data if there is no error in at least $\hat{k}$ encoded symbols.  Without loss
of generality, we assume that the data collector retrieves encoded symbols from
storage nodes $j_0$, $j_1$,$\ldots j_{\hat{k}-1}$. If no error is
present, the $\hat{k}$ symbols in the $i$-th group of any data node can be
recovered by solving the following system of linear equations:
\begin{eqnarray}
[u_{i0},u_{i1},\ldots,u_{i(\hat{k}-1)}]\hat{\G}=[c_{ij_0},c_{ij_1},\ldots,c_{ij_{\hat{k}-1}}]~,\label{uGc}
\end{eqnarray}
where
$$
\hat{\G}=\left[\begin{array}{cccc} 1&1&\cdots&1\\
\alpha^{j_0}&\alpha^{j_1}&\cdots&\alpha^{j_{\hat{k}-1}}\\
(\alpha^{j_0})^2&(\alpha^{j_1})^2&\cdots&(\alpha^{j_{\hat{k}-1}})^2\\
&&\vdots&\\
(\alpha^{j_0})^{\hat{k}-1}&(\alpha^{j_1})^{\hat{k}-1}&\cdots&(\alpha^{j_{\hat{k}-1}})^{\hat{k}-1}
\end{array}\right]~.$$ $\hat{\G}$ can be constructed by the primitive element
and the index associated with $c_{ij_d}$, $0\le d\le \hat{k}-1$.  

When the number of erroneous (or compromised) nodes is unknown but is bounded, the
proposed progressive  procedure for data retrieval
minimizes communication cost without any prior knowledge regarding
failure models of nodes.

From Section~\ref{sect:related}, we know that RS codes can recover from any $v$ errors if
$v \le \lfloor\frac{n-\hat{k}-s}{2}\rfloor$. Therefore, if the number of compromised
nodes ($v$) is small, more erasures ($s$) can be tolerated, and less nodes need
to be accessed (by treating them as unavailable). The data retrieval procedure
proceeds in stages. At stage $l$, $l$ errors are assumed. If RS decoding fails
or the decoded information symbols fail the CRC check, there must exist more
erroneous nodes than RS error-erasure decoding can handle at this stage. In
order to correct {\it one} more error, {\it two} more symbols need to be collected,
since the number of erasures allowed is reduced by two. Therefore, the total
number of symbols retrieved at stage $l$ is $\hat{k}+2l$. 

This procedure is clearly optimal in communication costs as additional symbols
are retrieved only when necessary. However, if applied naively, its computation
cost can be quite high since RS decoding shall be performed at each stage.  For
example, when $n = 1023$, $\hat{k} = 401$, with 1\% error probability defined as
probability that a storage node is faulty, our analytical results
from Section~\ref{sect:complexity} show that--on average-$409.2$ storage nodes need
to be accessed.  That is, the decoding needs to be done $\left\lceil\frac{409.2-401}{2}\right\rceil=5$ times.
On the other hand, consider a naive scheme that
retrieves coded symbols from each of $n$ storage nodes and decodes only once.
The naive scheme may incur less computation, but suffers from a high
communication cost. Such trade-offs between computation and
communication are avoidable as we show in Section~\ref{sect:algo},
where  we devise an algorithm that
can utilize intermediate computation results from previous stages and perform
RS decoding incrementally. Combined with the incremental decoding of stored
symbols, the proposed progressive data retrieval scheme (detailed in
Algorithm~\ref{algo:retrieval}) is both computation and communication
efficient. For simplicity, Algorithm~\ref{algo:retrieval} is  presented only
for one group of encoded symbols. It is applied to all groups of encoded
symbols to retrieve all the original data.

\begin{algorithm}[h]
\Begin {
$i \leftarrow \hat{k}$;

The data collector randomly chooses $\hat{k}$ storage nodes and retrieves encoded data,
$\ccc_i = [c_{j_0},c_{j_1},\ldots,c_{j_{\hat{k}-1}}]$;\\ 
$\rr_i = \ccc_i$

\Repeat {$i \ge n-1$} {
$\uu = \rr_i\hat{G}^{-1}$;

\lnl{crc} \eIf{$CRCTest(\uu) = SUCCESS$} {
Delete CRC checksum from $\uu$ to obtain $\uu_0$; \\
\Return $\uu_0$;
} {
\Repeat {\lnl{rs} \{$(\rr_i = IRD(\ccc_i)) = SUCCESS$ $\parallel$ $i \ge n - 1$\}} {
$i \leftarrow i + 2$

Two more encoded data from remaining nodes $i_1, i_2$, are retrieved

$\ccc_i \leftarrow \ccc_{i-2} \cup \{c_{i_1}, c_{i_2}\}$

} 
}
}

\Return FAIL;
}
\caption{Progressive Data Retrieval}
\label{algo:retrieval}
\end{algorithm}

In Algorithm~\ref{algo:retrieval}, for each $i$ (or accordingly stage $l =
(i-\hat{k})/2$ where the number of errors $v > l$), the decoding process declares
errors in one of two cases. In Line~\ref{rs}, the proposed incremental RS decoding
algorithm ($IRD()$) may fail to produce decoded symbols. Otherwise, in
Line~\ref{crc}, the decoded symbols fail the CRC check. Our implementation
(Section~\ref{sect:eval}) shows that the former happens frequently. Thus, in
most cases, CRC checking is carried out only once throughout the entire
decoding process.
\section{Progressive Decoding}
\label{sect:algo}
In this section, we present the incremental RS decoding algorithm. Compared to
the classic RS decoding, it utilizes intermediate computation results and
decodes incrementally as more symbols become available. 
\subsection{The basic algorithm}
Given the received coded symbols $[r_0, r_1, \ldots, r_n]$ with erasures
set to be zero, the generalized syndrome polynomial $S(x)$ can be
calculated~\cite{ARA92} as follows:
\begin{align}
S(x)=\displaystyle\sum_{j=0}^{n-1}r_j\alpha^{jb}\frac{T(x)-T(\alpha^j)}{x-\alpha^j}
=\displaystyle\sum_{j=0}^{n-1}\lambda_j\alpha^{jb}\frac{T(x)-T(\alpha^j)}{x-\alpha^j}~,
\label{syndrome}
\end{align}
where $T(x)$ is an arbitrary polynomial with degree $(n-\hat{k})$. Assume that $v$ errors occur in unknown locations
$j_1,j_2,\ldots,j_v$ and $s$ erasures in known locations $m_1,m_2,\ldots, m_s$ of the received polynomial. Then
$$e(x)=e_{j_1}x^{j_1}+e_{j_2}x^{j_2}+\cdots+e_{j_v}x^{j_v}$$
and
$$\gamma(x)=\gamma_{m_1}x^{m_1}+\gamma_{m_2}x^{m_2}+\cdots+\gamma_{m_s}x^{m_s}~,$$
where $e_{j_\ell}$ is the value of the $\ell$-th error, $\ell=1, \cdots, v$, and $\gamma_{m_\ell}$ is the value of the $\ell$-th erasure, $\ell=1, \cdots, s$. Since the received values in the erased positions are zero, $\gamma_{m_\ell}=-c_{m_\ell}$ for $\ell= 1, \cdots, s$. The decoding
process is to find all $j_\ell$, $e_{j_\ell}$, $m_\ell$, and $\gamma_{m_\ell}$. Let ${\bf E}=\{j_1,\cdots,j_v\}$, ${\bf M}=\{m_1,\cdots,m_s\}$, and ${\bf D}={\bf E}\cup {\bf M}$. Clearly, ${\bf E}\cap {\bf M}=\emptyset$.
It has been shown that a key equation for decoding is
\begin{eqnarray}
\Lambda(x)S(x)=\Psi(x)T(x)+\Omega(x)~,\label{key_equation}
\end{eqnarray}
where
\begin{eqnarray}
\Lambda(x)&=&\prod_{j\in {\bf D}}(x-\alpha^j)=\prod_{j\in {\bf E}}\left(x-\alpha^j\right)\prod_{j\in {\bf M}}\left(x-\alpha^j\right)\nn\\
&=&\Lambda_{\bf E}(x)\Lambda_{\bf M}(x)\label{sigma}\\
\Psi(x)&=&\sum_{j\in {\bf D}}\lambda_j\alpha^{jb}\prod_{{i\in {\bf D}}\atop{i\neq j}}\left(x-\alpha^i\right)\label{psi}\\
\Omega(x)&=&-\sum_{j\in {\bf D}}\lambda_j\alpha^{jb}T(\alpha^j)\prod_{{i\in {\bf D}}\atop{i\neq j}}\left(x-\alpha^i\right)~.\label{omega}
\end{eqnarray}
If $2v+s\le n-\hat{k}+1$, then~\eqref{key_equation} has a unique solution $\{\Lambda(x),\Psi(x),\Omega(x)\}$. Instead of solving \eqref{key_equation} by either the Euclidean or Berlekamp-Massey algorithm we introduce a
reduced key equation ~\cite{ARA92} that can be solved by the Welch-Berlekamp
(W-B) algorithm~\cite{MOO05}. It will be demonstrated that by using W-B
algorithm and the reduced key equation, the complexity of decoding can be reduced drastically. Let ${\bf T}=\{j|T(\alpha^j)=0\}$.  Let a set of
coordinates ${\bf U}\subset \{0,1,\ldots,n-1\}$ be defined by ${\bf U} ={\bf M}\cap {\bf T}$. A polynomial $\Lambda_{{\bf U}}(x)$ is then defined by
$\Lambda_{{\bf U}}(x)=\prod_{j\in {\bf U}}\left(x-\alpha^j\right),$ which is known for the
receiver since $T(x)$ and ${\bf M}$ are both known. Since $\Lambda_{\bf U}(x)$
divides both $\Lambda(x)$ and $T(x)$, according to~\eqref{key_equation}, it
also divides $\Omega(x)$. Hence, we have the following reduced key equation:
\begin{eqnarray}
\tilde\Lambda(x)S(x)=\Psi(x)\tilde{T}(x)+\tilde\Omega(x)~,\label{reduced_key_equation}
\end{eqnarray}
where
\begin{eqnarray*}
\Lambda(x)&=&\tilde\Lambda(x)\Lambda_{{\bf U}}(x)\\
T(x)&=&\tilde{T}(x)\Lambda_{{\bf U}}(x)\\
\Omega(x)&=&\tilde\Omega(x)\Lambda_{{\bf U}}(x)~.
\end{eqnarray*}
Note that $\tilde\Lambda(x)$ is still a multiple of the error location polynomial $\Lambda_{{\bf E}}(x)$. The reduced key equation can have a unique solution if
\begin{eqnarray}
\deg(\tilde\Omega(x))<\deg(\tilde\Lambda(x))<\frac{n-\hat{k}+1+s}{2}-|{\bf U}|~,\label{criterion}
\end{eqnarray}
where $\deg(\cdot)$ is the degree of a polynomial and $|{\bf U}|$ is the number of elements in set ${\bf U}$.

For all $j\in {\bf T}\backslash
{\bf U}$, by~\eqref{reduced_key_equation}, we have
\begin{eqnarray}\tilde\Lambda(\alpha^j)S(\alpha^j)=\tilde\Omega(\alpha^j)\label{W-B_key_equation}\end{eqnarray}
since $\tilde{T}(\alpha^j)=0$. Note that $\alpha^j$ is a sampling point and
$S(\alpha^j)$ the sampled value for~\eqref{W-B_key_equation}. The unique
solution $\{\tilde\Lambda(x),\tilde\Omega(x)\}$ can then be found by the W-B
algorithm with time complexity $O((n-\hat{k}-|{\bf U}|)^2)$~\cite{MOO05}.  Once all
coefficients of the errata polynomial are found, the error locations $j_\ell$
can be determined by successive substitution through Chien search~\cite{LIN04}.
When the solution of~\eqref{reduced_key_equation} is obtained, the errata
values can be calculated. Since there is no need to recover the errata values in our application
we omit the calculations. In summary, there are three steps in the
decoding of RS codes that must be implemented. First, the sampled values of
$S(\alpha^j)$ for $j\in {\bf T}\backslash {\bf U}$ must be calculated. Second, the W-B
algorithm is performed based on the pairs $\left(\alpha^j, S(\alpha^j)\right)$
in order to obtain a valid $\tilde\Lambda(x)$. If a valid $\tilde\Lambda(x)$ is
obtained, then error locations are found by Chien search; otherwise, decoding
failure is reported. 
\subsection{Incremental computation of $S(x)$, $\tilde\Lambda(x)$, $\tilde\Omega(x)$}
Let us choose
$$T(x)=\left(x-\alpha^{m_0}\right)\left(x-\alpha^{m_1}\right)\cdots\left(x-\alpha^{m_{n-\hat{k}-1}}\right)~,$$
where $m_\ell$ are those corresponding positions of missing data symbols after
the data collector has retrieved encoded symbols from $\hat{k}$ storage nodes. In the
decoding process, these are erased positions before the first iteration of
error-erasure decoding. Let ${\bf U}_0=\{m_0,\ldots,m_{n-\hat{k}-1}\}$.
The generator polynomial of the RS code encoded by~\eqref{eq:generator}
has $\alpha^{n-\hat{k}},\alpha^{n-\hat{k}-1},\ldots, \alpha$ as roots.
The error-erasure decoding algorithm is mainly based on W-B algorithm which is
an iterative rational interpolation method.

In the $\ell$-th iteration, $\ell$ errors are assumed in the data and the
number of erasures is $n-\hat{k}-2\ell$. Let $(j^{(\ell)}_1+1)$ and $(j^{(\ell)}_2+1)$ be the two storage nodes just accessed in the $\ell$-th iteration. Let ${\bf U}_\ell={\bf U}_{\ell-1}\backslash \{j^{(\ell)}_1,j^{(\ell)}_2\}$. Based on ${\bf U}_\ell$ the W-B algorithm will find $\tilde\Lambda^{(\ell)}(x)$ and $\tilde\Omega^{(\ell)}(x)$ which satisfy
$$\tilde\Lambda^{(\ell)}(\alpha^j)S^{(\ell)}(\alpha^j)=\tilde\Omega^{(\ell)}(\alpha^j)\mbox{ for all }j\in {\bf U}_0\backslash {\bf U}_{\ell}~,$$ where $S^{(\ell)}(x)$ is the generalized syndrome with $r_i=0$ for all $r_i\in {\bf U}_\ell$. It has been shown that $\deg(\tilde\Lambda^{(\ell)}(x))> \deg(\tilde\Omega^{(\ell)}(x))$ for any $\ell$ by a property of W-B algorithm. Thus, if $\deg(\tilde\Lambda^{(\ell)}(x))<\frac{n-\hat{k}+1+|{\bf U}_\ell|}{2}-|{\bf U}_\ell|=\ell+1/2$, then the unique solution will exist due to~\eqref{criterion}. By the definition of generalized syndrome polynomial in~\eqref{syndrome}, for $i\in {\bf U}_0\backslash {\bf U}_\ell$, we have

\begin{eqnarray}
S^{(\ell)}(\alpha^i)&=&\sum_{j=0}^{n-1}r_j\alpha^j\frac{T(\alpha^i)-T(\alpha^j)}{\alpha^i-\alpha^j}\nn\\
&=&\sum_{j=0\atop j\notin {\bf U}_0}^{n-1}r_j\alpha^j\frac{T(\alpha^j)}{\alpha^j-\alpha^i}+r_i\alpha^iT^\prime(\alpha^i)\nn\\
&=&\sum_{j=0\atop j\notin {\bf U}_0}^{n-1}\frac{F_j}{\alpha^j-\alpha^i}+r_i\alpha^iT^\prime(\alpha^i)~,\label{syndrome_j}
\end{eqnarray}
where $T^\prime(x)$ is the derivative of $T(x)$ and $F_j=r_j\alpha^jT(\alpha^j)$. Note that $T^\prime(\alpha^i)=\prod_{j\in {\bf U}_0\atop m_j\neq i}\left(\alpha^i-\alpha^{m_j}\right).$ It is easy to see that $S^{(\ell)}(\alpha^i)$ is not related to any $r_j$, where $j\in {\bf U}_0$ and $j\neq i$. Hence, $S^{(\ell-1)}(\alpha^i)=S^{(\ell)}(\alpha^i)$ for all $i\in {\bf U}_0\backslash {\bf U}_{\ell-1}$. This fact implies that all sampled values in previous iterations can be directly used in current iteration of the W-B algorithm.

Define $\mbox{rank}[N(x),W(x)]=\max[2\deg(W(x)),1+2\deg(N(x))].$ The
incremental RS decoding algorithm is described in
Algorithm~\ref{algo:incremental}. Upon success, the incremental RS decoding
algorithm returns $\hat{k}$ non-error symbols. The procedure will report failure either as the result
of mismatched degree of the error locator polynomial, or insufficient number of
roots found by Chien search (Line~\ref{chien}). In both cases, no further erasure decoding is
required.  This reduces the decoding computation time.
\begin{algorithm}[h]
\SetLine
\SetKwInOut{Initialization}{init}
\SetKwInOut{Input}{input}
\SetKwInOut{Output}{output}
\Initialization{Calculate $F_j$ given in~\eqref{syndrome_j} for all $j\notin {\bf U}_0$. \\
$\ell\leftarrow 0$; $\tilde\Lambda^{(0)}(x)\leftarrow 1$; \\ $\tilde\Omega^{(0)}(x)\leftarrow 0$; $\Phi^{(0)}(x)\leftarrow 0,\Theta^{(0)}(x)\leftarrow 1$.}
\Input{stage $l$, two new symbols at storage nodes $(j_1^{(\ell)}+1)$ and $(j_2^{(\ell)}+1)$}
\Output{FAIL or non-error symbols $\rr$}
\Begin{

\ForEach{$i=1,2$} {
\lnl{error-syndrome} $x^{(\ell)}_i\leftarrow \alpha^{j^{(\ell)}_i}$ and $y^{(\ell)}_i\leftarrow S^{(\ell)}(x^{(\ell)}_i)$}

\For{$i=1$ {\bf to} $2$}{ 
$b^{(\ell-1)}_i\leftarrow \tilde\Omega^{(\ell-1)}(x^{(\ell)}_i)-y^{(\ell)}_i\tilde\Lambda^{(\ell-1)}(x^{(\ell)}_i)$;

\eIf {$b^{(\ell-1)}_i=0$} {
$\tilde\Lambda^T(x)\leftarrow \tilde\Lambda^{(\ell-1)}(x)$;
$\tilde\Omega^T(x)\leftarrow \tilde\Omega^{(\ell-1)}(x)$;
$\Theta^{T}(x)\leftarrow (x-x^{(\ell)}_i)\Theta^{(\ell-1)}(x)$;
$\Phi^{T}(x)\leftarrow (x-x^{(\ell)}_i)\Phi^{(\ell-1)}(x)$ } {
$a^{(\ell-1)}_i\leftarrow \Theta^{(\ell-1)}(x^{(\ell)}_i)-y^{(\ell)}_i\Phi^{(\ell-1)}(x^{(\ell)}_i)$;
$\Theta^{T}(x)\leftarrow (x-x^{(\ell)}_i)\tilde\Omega^{(\ell-1)}(x)$;
$\Phi^{T}(x)\leftarrow (x-x^{(\ell)}_i)\tilde\Lambda^{(\ell-1)}(x)$; 
$\tilde\Omega^T(x)\leftarrow b^{(\ell-1)}_i\Theta^{(\ell-1)}(x)-a^{(\ell-1)}_i\tilde\Omega^{(\ell-1)}(x)$; 
$\tilde\Lambda^T(x)\leftarrow b^{(\ell-1)}_i\Phi^{(\ell-1)}(x)-a^{(\ell-1)}_i\tilde\Lambda^{(\ell-1)}(x)$.
}

\If{$\mbox{rank}[\tilde\Omega^T(x),\tilde\Lambda^T(x)]>\mbox{rank}[\Theta^{T}(x),\Phi^{T}(x)]$}{
swap $[\tilde\Omega^T(x),\tilde\Lambda^T(x)]\leftrightarrow [\Theta^{T}(x),\Phi^{T}(x)]$.
}

\eIf{$i=1$}{
$\tilde\Omega^{(\ell-1)}(x)\leftarrow \tilde\Omega^T(x)$; 
$\tilde\Lambda^{(\ell-1)}(x)\leftarrow \tilde\Omega^T(x)$; 
$\Theta^{(\ell-1)}(x)\leftarrow \Theta^T(x)$, $\Phi^{(\ell-1)}(x)\leftarrow \Phi^T(x)$; 
} {
$\tilde\Omega^{(\ell)}(x)\leftarrow \tilde\Omega^T(x)$; 
$\tilde\Lambda^{(\ell)}(x)\leftarrow \tilde\Omega^T(x)$; 
$\Theta^{(\ell)}(x)\leftarrow \Theta^T(x)$; 
$\Phi^{(\ell)}(x)\leftarrow \Phi^T(x)$.
}
}
\If {$\deg(\tilde\Lambda^{(\ell)}(x))\neq \ell$}{
\Return FAIL;
}

NumErrorLoc = ChienSearch($\tilde\Lambda^{(\ell)}(x)$). 

\lnl{chien} \If{$NumErrorLoc > n - \hat{k} \parallel NumErrorLoc \neq \deg(\tilde\Lambda^{(\ell)}(x))$} {
\Return FAIL;
}
\Return $\hat{k}$ non-error symbols $\rr$;
}
\label{algo:incremental}
\caption{Incremental RS Decoding $IRD$}
\end{algorithm}
%
\section{Complexity Analysis}
\label{sect:complexity}
This section provides a complexity analysis for data storage and retrieval in Sections~\ref{sect:enc_comm}
and~\ref{sect:retrieve} respectively. Both have computational and communication costs associated with them. Specifically, data
storage is composed of both an encoding and data dissemination phase, while data retrieval is composed of both an incremental collection
and decoding phase.
Included in 
Section~\ref{sect:retrieve} are Monte-Carlo simulations that are consistent with our data retrieval complexity analysis.
 Finally, in Section~\ref{sect:dynamic}, the benefit of relaxing the
$\hat{k}=k$ constraint that was imposed by our previous work~\cite{Han10-Infocom}, is provided.

\subsection{Data Storage}
\label{sect:enc_comm}
From Section~\ref{sect:data_storage}, the  communication cost incurred by the encoded data generated by a
data node is $nm\left\lceil\frac{\lceil T/m\rceil}{\hat{k}}\right\rceil$ bits. The total communication cost 
is then a factor of $k$ more.
Also, it is easy to see that the total bits stored in each storage node is
$km\left\lceil\frac{\lceil T/m\rceil}{\hat{k}}\right\rceil$, which is approximately
$T$ when $\hat{k}=k$ and $T$ is much larger than $mk$. 
Assuming a software implementation on field operations without use of look-up tables,
the computation complexity of encoding can be estimated as follows. 
Given that computation of one multiplication in $GF(2^m)$ is of $m^2$ bit
exclusive-ORs. At the data node, $\hat{k}n\left\lceil\frac{\lceil
T/m\rceil}{\hat{k}}\right\rceil$ multiplications are performed, which is equivalent
to $\hat{k}n\left\lceil\frac{\lceil T/m\rceil}{\hat{k}}\right\rceil m^2$ bit exclusive-ORs.
\subsection{Data Retrieval}
\label{sect:retrieve}
Given a set of coded symbols, Section~\ref{sect:decoding_complexity}
analyzes the computational decoding costs. Then a derivation of the 
communication complexity is provided in Section~\ref{sect:analysis}
\subsubsection{Decoding}
\label{sect:decoding_complexity}
In the subsequent complexity analysis, the worst case is assumed, namely, no
failure on decoding is reported in Algorithm~\ref{algo:incremental}
(Line~\ref{chien}), and the algorithm runs to completion.

In CRC checking, one polynomial division is performed. Since the dividend is of
degree $T-1$ and the divider is of degree $r$, the computation complexity is $O(Tr)$.

Let $v$ be the number of errors when the decoding procedure is completed.
In the $\ell$-th iteration, $\ell$ errors are assumed in the data and the
number of erasures is $n-\hat{k}-2\ell$. We first need to calculate two
syndrome values. This can be obtained by the $F_j$ calculated initially. For instance, in the first iteration, according to~\eqref{syndrome_j}, the
computation complexity is of $O(\hat{k}(n-\hat{k}))$ since there are $\hat{k}$ $F_j$'s to be calculated and each is a product of $n-\hat{k}$ terms. In the next iteration, two more symbols are
added to~\eqref{syndrome}. Hence, the updated syndrome values can be obtained by an extra
$O(\hat{k})+O(n-\hat{k})$ computations. To find the error-locator polynomial, the W-B
algorithm is performed two steps in each iteration with complexity $O(\ell)$.
Since we only consider software implementation,  the Chien search can be
replaced by substituting a power of $\alpha$ into the error-locator polynomial.
It needs to test for at most $\hat{k}+\ell$ positions to locate $\hat{k}$ non-error
positions such that it takes $O((\hat{k}+\ell)\ell)$ computations. Finally, inversion
of a Vandermonde matrix $\hat{G}$ can be done in $O(\hat{k}\log^2\hat{k})$~\cite{GOH94}, though for implementation purposes,
we use a $O(\hat{k}^2)$ inversion algorithm (see, e.g.,~\cite{william1988numerical}).
In
summary, the computation in the $\ell$-th iteration for $\ell>1$ is
\begin{eqnarray*}
L_v(\ell)&=&
O(\hat{k}^2)+O(n-\hat{k})+O(\hat{k}\ell+\ell^2)~.
\end{eqnarray*}
Counting for $v$ iterations and the complexity of calculating $F_j$ we have
\begin{eqnarray}
L_v
&=&O(v\hat{k}^2)+O(\hat{k}(n-\hat{k}))+O(v^2\hat{k})\nonumber\\
&&+O(v(n-\hat{k}))+O(v^3)~.\label{eq:dec_comp}
\end{eqnarray}
Note that the computation complexity is measured by finite field multiplications,
which is  equivalent to $m^2$ bit exclusive-ORs.  Since the correctable number
of errors $v$ is at most $(n-\hat{k})/2$, the decoding complexity is at most
$O(\hat{k}(n-\hat{k})^2)$. For small $v$, 
the second term $O(\hat{k}(n-\hat{k}))$ dominates, which corresponds to syndrome computation. 
Note that the decoding procedure needs to be performed $k\left\lceil\frac{\lceil
T/m\rceil}{\hat{k}}\right\rceil$ times in order to decode all data.
%
\subsubsection{Communication}
\label{sect:analysis}
In this section, we provide a probabilistic analysis of the cost of
communication by determining the number of stages the algorithm needs to take,
and the probability of successful execution. Given $n$ storage nodes and a
$(n,\hat{k})-$MDS code, the minimum and maximum number of storage nodes to access
in the proposed scheme is $\hat{k}$ and $n$ respectively. We assume that
the CRC code always detect an error if it occurs.  Without loss of
generality, we assume that all failures are Byzantine failures, since $s$ crash-stop
failures can be easily modeled by replacing $n$ with $n-s$. An important metric
of the decoding efficiency is the average number of accessed storage nodes when
the probability of compromising each storage node is $p$. Failure to recover
data correctly may occur in two cases. First, $v > n-\hat{k}$, i.e., there are
insufficient number of healthy storage nodes. Second,
$\lfloor\frac{n-\hat{k}}{2}\rfloor < v < n-\hat{k}$, in which the sequence of accessing
determines the outcome (success or failure) of the decoding process.  For
example, if the first $v$ nodes accessed are all compromised nodes, correct
decoding is impossible. In both cases, the decoding algorithm stops after $n$
accesses and declares a failure.  The communication cost is $n$. 
The main result is summarized in the following theorem.
\begin{theorem}
With the progressive data retrieval scheme, the average number of accesses as well as the 
probability of successful decoding are given in Eq.~\eqref{eq:average} and~\eqref{eq:sucprob} respectively.
\begin{table*}
\par\noindent\rule{0.9\textwidth}{0.4pt}%
\begin{eqnarray}
\nonumber
\bar N(n,\hat{k})&=&\sum_{v=0}^{n-\hat{k}}{n\choose v}p^v(1-p)^{n-v}\sum_{i=0}^{\min(v, \lfloor{\frac{n-\hat{k}}{2}}\rfloor, n-v-\hat{k})}(\hat{k}+2i)\frac{{n-v\choose i+\hat{k}-1}{v \choose i}}{{n \choose 2i+\hat{k}-1}}\times \frac{\hat{k}}{i+\hat{k}} \times \frac{n - v - (i+\hat{k}-1)}{n - (2i+\hat{k}
-1)} \\
\nonumber
&+& \sum_{v=0}^{n-\hat{k}}n{n\choose v}p^v(1-p)^{n-v}\left(1-\sum_{i=0}^{\min(v, \lfloor{\frac{n-\hat{k}}{2}}\rfloor, n-v-\hat{k})}\frac{{n-v\choose i+\hat{k}-1}{v \choose i}}{{n \choose 2i+\hat{k}-1}}\times \frac{\hat{k}}{i+\hat{k}} \times \frac{n - v - (i+\hat{k}-1)}{n - (2i+\hat{k}
-1)}\right) \\
&+ &\sum_{v=n-\hat{k}+1}^{n}n{n\choose v}p^v(1-p)^{n-v}~.
\label{eq:average}
\end{eqnarray}

\begin{equation}
\mbox{Pr}_{suc}(n,\hat{k})=\sum_{v=0}^{n-\hat{k}}{n\choose v}p^v(1-p)^{n-v}\sum_{i=0}^{\min(v, \lfloor{\frac{n-\hat{k}}{2}}\rfloor, n-v-\hat{k})}\frac{{n-v\choose i+\hat{k}-1}{e \choose i}}{{n \choose 2i+\hat{k}-1}}\times \frac{\hat{k}}{i+\hat{k}} \times \frac{n - v - (i+\hat{k}-1)}{n - (2i+\hat{k}
-1)}~.
\label{eq:sucprob}
\end{equation}
\par\noindent\rule{0.9\textwidth}{0.4pt}%
\end{table*}
\label{thm:average_access}
\end{theorem}
The details of the proof can be found in the Appendix, and numerical 
backing of this analysis is illustrated in what follows.

\paragraph*{Numerical Corroboration}
\label{sect:num}
\begin{figure}[thp]
\begin{center}
\includegraphics[width=2.5in]{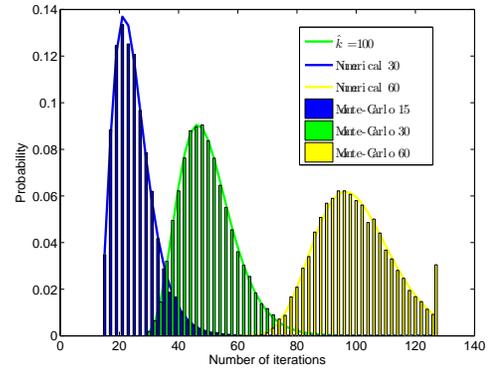}
\caption{The effect of $\hat{k}$ on the data retrieval cost for a  $(127,\hat{k})-$MDS; $k=30$. The error probability here is $p=0.2$; $\hat{k}\in\{\frac{k}{2},k,2k\}$. }
\label{fig:khat}
\end{center}
\end{figure}
We verify the correctness of the analytical model using Monte-Carlo simulations
implemented in Matlab. Figure~\ref{fig:khat} shows the distribution of the number of
storage nodes accessed when the algorithm terminates, and the number of iterations
correspond to the number of node accesses during data retrieval.
The bar charts depict 
histograms from Monte-Carlo simulations with $5000$ runs, and the curves
represent the numerical results from our analytical model. We choose $n = 127$, $k=30$ and $p =
0.2$ so that $5000$ runs give sufficient statistics in the simulations. From Figure~\ref{fig:khat},
it can be observed that the analytical results agree well with the Monte-Carlo simulations.
Note that the number of information symbols, $\hat{k}$, yields different distribution results.
Specifically, increasing $\hat{k}$ reduces storage--as derived in Section~\ref{sect:enc_comm}.
However, Figure~\ref{fig:khat} shows that the expense--in terms of data retrieval--is undesirably high.
\begin{figure}[thp]
\begin{center}
\includegraphics[width=2.5in]{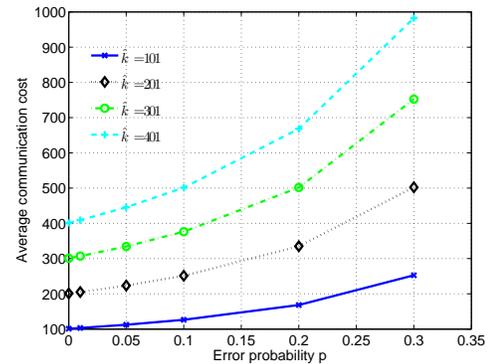} 
\caption{Number of storage nodes accessed as a function of the probability of malicious attacks for a $(1023,\hat{k})-$MDS; $k = 101$.}
\label{fig:ret}
\end{center}
\end{figure}

Next, we fix $n = 1023$, and vary the number of information symbols, $\hat{k}$, from
$101$ to $401$ and the error probability $p$ from $0$ to $0.3$, while keeping the number
of data nodes constant.
Figure~\ref{fig:ret} shows the increasing communication cost as the
probability of failures increases. The number of crash-stop failures is set to
zero, and all Byzantine failures result in incorrect data.  Clearly, when the
error probability $p$ is small, the communication cost is close to $\hat{k}$. And when $p$ increases, the
communication cost monotonically increases, as expected. We also analyze the success rate of decoding. And observe
that for $\hat{k}\in\{101,201,301\}$ and $p\in\{0,0.05,...,0.3\}$, decoding will always be successful. However, for $\hat{k}=401$, decoding is always successful \emph{only} for $p\in\{0,0.05,...,0.25\}$. When $p=0.3$, the probability of successful decoding is only $60\%$.

\subsection{The Dynamics of $\hat{k}$}
\label{sect:dynamic}
One advantage of the proposed scheme is that the number of information symbols, $\hat{k}$,
is not tied to the number of data nodes $k$.
Hence, one may choose appropriate values of $\hat{k}$ and $k$ for any given application. 
For example, in wireless sensor networks, data nodes
are power-limited but the given data collector typically has no power constraint.
In such applications, one should reduce the 
cost to disseminate coded symbols from data nodes to storage nodes,
and shift the cost to the data collection phase. This can be done 
by choosing $\hat{k} > k$. As shown in Section~\ref{sect:enc_comm}, the total 
communication cost is
$$knm\left\lceil\frac{\lceil T/m\rceil}{\hat{k}}\right\rceil$$ such that this cost roughly varies 
linearly with the ratio of $k$ and $\hat{k}$. If one takes $\hat{k}=2k$, then 
this dissemination cost is half the dissemination cost of $\hat{k}=k$. However, from our 
analytical model, we see that the data retrieval cost is then doubled. 
\section{Relative Analysis}
\label{sect:era_comp}
In this section, we compare the proposed scheme (IRD) to Decentralized Erasure Codes
(DEC) proposed by Dimakis et al.~\cite{DIM06}, and Decentralized Fountain Codes (DFC) proposed by Lin et al.~\cite{lin2007dpl}. To make this comparison, we assume
that there are $k$ data nodes that contain
the data to be redundantly stored amongst $n$ storage nodes.  The $k$ data
 nodes collectively generate $kT_0$ bits, where there are $T_0$ bits per node. As mentioned in Section~\ref{sect:data_storage}, each IRD data node adds $r$ CRC bits to its data. For brevity, we therefore will write $T = T_0 + r$ for the data size of each IRD data node. To facilitate understanding,
we set $\hat{k}=k$ in all schemes utilizing information symbols.

In this analysis, data  and storage nodes' data can be partitioned into \emph{data} and \emph{storage} symbols respectively. The number of bits in a data symbol is always $m$. Depending on the storage scheme, a storage symbol may be larger than $m$. Multiplication of a $m_1$-bit and $m_2$-bit symbol, $m_1\ge m_2$, requires $m_1m_2$ XORs, while an addition requires only $m_1$, where the field of operation for the symbols is $\mathbb{F}_{m_1}$. 

Similar to Section~\ref{sect:data_storage}, encoding and decoding for all schemes is done in groups. DEC and DFC both have $1$ data symbol per data node per group, while IRD encodes $k$ data symbols per data node per group. Consequently, each data node in DEC/DFC and IRD has $\frac{T_0}{m}$ and $\frac{T}{mk}$ groups per data node, respectively. As we will see, although fountain coding minimizes the encoding and decoding complexities, IRD minimizes 
communication significantly, especially for decoding.

Quantifying the performance over storage codes requires a comparison over the metrics shown in Table~\ref{tab:comparison}.
Respectively, these metrics include the $n$ storage nodes' redundancy--total bits stored; storage nodes' overhead required for decoding; dissemination cost--communication cost between data and storage nodes; collection cost for $1/k$ and all of the original data, respectively; encoding computation cost; a data-collector's decoding computation cost; ability to detect and correct errors; and finally, ability to deterministically guarantee reconstruction of the original data.
We now provide a qualitative comparison amongst three storage schemes, based on these metrics.
\begin{table*}[tbp]
\caption{Performance comparison of erasure coding schemes for storage}
\label{tab:comparison}
\begin{center}
\begin{tabular}{|c|c|c|c|}
\hline
& DEC & DFC & IRD\\  
\hline
\hline
Storage & $\frac{nT_0\log q}{m}$ & $nT_0$ &  $nT$\\
Overhead & $\frac{nT_0}{m}\log k(\log q + \log k + \log\frac{T_0}{mk})$ & $\frac{nT_0}{m}\cdot\log\frac{k}{\delta}\cdot(\log k + \log\frac{T_0}{mk})$ & $\frac{nT}{m}\left(\log k + \log\frac{T}{mk^2}\right)$\\
\hline
Dissemination & $nT_0\log k$ & $nT_0\log\frac{k}{\delta}$ & $nT$\\
$1$-collection &  $\frac{kT_0\log q}{m}$ &  $T_0\left(k+\sqrt k\log^2\frac{k}{\delta}\right)$ & $T$\\
$k$-collection &  $\frac{kT_0\log q}{m}$ &  $T_0\left(k+\sqrt k\log^2\frac{k}{\delta}\right)$ & $kT$\\
\hline
Encoding & $nT_0\log (q+k)$ & $nT_0\log\frac{k}{\delta}$ & $nkTm$\\ 
Decoding & $\frac{k^3T_0\log^2 q}{m}$ & $kT_0\log\frac{k}{\delta}$ & $k^2Tm$\\
\hline
Error detection & no & no & yes\\
Error correction & no & no & yes\\
Deterministic guarantee & no & no & yes\\
\hline
\end{tabular}
\end{center}
\end{table*}

\subsection{Decentralized erasure codes}
DEC have been applied in wireless or wired networks, where a data collector accesses {\it any} $k$ out of $n$
storage nodes for data reconstruction.
Each
storage node selects random and independent coefficients in a finite field
$\mathbb{F}_{q}$, and stores a linear combination of all the received data (modulo
$q$). Randomized linear codes are used, where each
data node routes its packet multiple times to at least $\frac{n}{k}\log k$ storage nodes, so that the $n$ storage nodes collectively store $\left(\frac{n}{k}\log k\right) kT_0$ bits.

The storage complexity per storage node is $\frac{T_0}{m}\cdot\log q$ bits, which holds because i) arithmetic is done in $\mathbb{F}_q$, which means each stored symbol has $\log q$ bits, and ii) there are $\frac{T_0}{m}$ groups.

The overhead can be calculated as follows. Since each storage node stores the linear combination coefficients and there are  $\log k$ data nodes connected to a storage node, the overhead--to store the coefficients--per storage node is 
$\log k(\log q+\log k + \log\frac{T_0}{m})$ bits. For any storage node, note that the $\log k$ outside the parentheses denotes the number of nodes connecting to the storage node, while the $\log k$ term inside the parentheses are the bits to identify any connecting data node~\cite{DIM06}. The last parenthesized term identifies the coding group.

The DEC dissemination cost is given by $$k\cdot \frac{n}{k}\log k\cdot\frac{T_0}{m}\cdot m\text{ bits}$$  because i) there are $k$ data  nodes, ii) each data node repeatedly sends its data out $\frac{n}{k}\log k$ times, and iii) there are $\frac{T_0}{m}$ groups.

Since the code structure is inherently random, $k$ nodes must be contacted in order to obtain any one symbol. Specifically, $k$ symbols are collected from $k$ storage nodes to reconstruct the generated data, per group.

The encoding cost per storage node is given by: 
$$(m\log q)\cdot\log k\cdot\frac{T_0}{m}\text{ XORs}$$
because i) the cost of a linear combination (multiplication) is $m\log q$ bit operations, since a combination coefficient and a data symbol are $\log q$ and $m$ bits respectively, ii) a storage symbol is the result of $\log k$ linear combinations, and iii) there are $\frac{T_0}{m}$ groups to be encoded.
Similarly, the decoding complexity to reconstruct the entire data is given by: 
$$(\log^2 q)\cdot k^3\cdot\frac{T_0}{m}\text{ XORs.}$$
This complexity can be derived from the following: i) a multiplication costs $\log^2 q$ bit operations, since each storage node stores $q$-bit symbols, ii) matrix inversion is performed, and iii) there are $\frac{T_0}{m}$ groups to be decoded.

Although DEC can
be efficiently constructed, error-detection and correction are both infeasible: Assuming the use of CRC for error-detection, a data-collector must continue to enumerate all possible $k$ symbols from $k$ out of $n$ storage nodes, until the original data can be reconstructed correctly. Therefore, DEC cannot be applied to dependable storage systems where data integrity is desired in the midst of errors and malicious users.

\subsection{Decentralized fountain codes}
DFC is a decentralized LT code~\cite{luby2002lt}, and were proposed for the special purpose of guaranteeing data availability in the presence of crash-stop failures for networks with several data generators and storage nodes. Storage node, $s_i$, where $i=1,...,n$, chooses a \emph{degree}, $d_i$, which is defined as the number of data symbols from which to form a linear combination. $s_i$ then linearly combines $d_i$ data symbols--using the XOR operation--from $d_i$ arbitrarily chosen
data generators. DFC--like other fountain codes--trades communication for computation: decoding requires more than $k$ storage nodes to be contacted, though both encoding and decoding computations are linear in the number of original symbols. In performance evaluation, we assume that $1-\delta$ is the probability of successful decoding. 
Different from DEC, instead of pulling data from candidate data  nodes, a deterministic and probabilistic scheme is devised to push data from the
source nodes to storage nodes~\cite{lin2007dpl}. Aside from using fountain codes, the authors use \emph{random walks} to remove the need for a geometric routing protocol for propagating data from data  nodes to storage nodes.

The storage complexity per storage node is
$\frac{T_0}{m}\cdot m$ bits because there are $\frac{T_0}{m}$ groups and $m$ bits per storage symbol. Note that--unlike DEC--the number of bits in a storage symbol is independent of the size of the operating field.

The overhead complexity per storage node is given by $$\frac{T_0}{m}\cdot\log\frac{k}{\delta}\cdot(\log k + \log\frac{T_0}{m})\text{ bits.}$$ The derivation here is similar to that of DEC with the following differences: the average degree of a storage symbol is $\log\frac{k}{\delta}$ and there are \emph{no linear combination coefficients}, since every linear combination is simply an XOR of a set of data symbols. 

The dissemination cost for DFC is given by
$$n\cdot\log\frac{k}{\delta}\cdot\frac{T_0}{m}\cdot m\text{ bits.}$$ This holds because i) there are $n$ storage nodes,
and ii) each node stores $\log\frac{k}{\delta}$ symbols on average.

DFC is an LT code that is not \emph{systematic}. Also, to reconstruct any data, more than $k$ nodes must be contacted. Specifically, the communication cost to collect all data symbols is given by
$$T_0\left(1+\frac{\log^2\frac{k}{\delta}} {\sqrt k}\right)\text{ bits.}$$ This holds because $k+\sqrt k\log^2\frac{k}{\delta}$ symbols must be collected for successful data reconstruction of any $k$ symbols. 

Encoding and decoding can be very efficient in DFC. The encoding complexity per storage node is given by
$$m\cdot\log\frac{k}{\delta}\cdot\frac{T_0}{m}\text{ XORs}$$ since i) the cost of an XOR of two $m$-bit symbols is $m$ XORs and
ii) each encoded symbol is the XOR of $\log\frac{k}{\delta}$ $m$-bit data symbols, where $\log\frac{k}{\delta}$ is the average degree of an encoded symbol.
In a similar manner, the decoding for DFC codes is given by
$$\frac{T_0}{m}\cdot k\log\frac{k}{\delta}\cdot m\text{ XORs}$$ because DFC uses the $k\log\frac{k}{\delta}$ belief propagation algorithm for decoding.
DFC is most efficient in decoding. However, like all fountain codes, the decoding efficiency comes at a communication trade-off that is determined by the parameter, $\delta$.

\subsection{IRD}
Each data  node in IRD encodes its own data, with $k$ symbols per group to the $n$ storage nodes. Therefore, altogether there are 
$\frac{T}{mk}$ groups. Note that IRD has the same storage complexity as DFC even though the numbers of groups and data symbols per group differ. Because IRD utilizes a code structure known to all storage nodes, it has the minimum overhead complexity per storage node:
$$\frac{T}{mk}\cdot k\left(\log k + \log\frac{T}{mk}\right)\text{ bits.}$$

Unlike DEC and DFC, IRD does not replicate transmissions to storage nodes, and therefore its dissemination cost exactly equals its storage complexity, leaving IRD with the minimum dissemination cost as well.  
IRD is also preferable because it is systematic, allowing partial collection of a subset of data. Since $k$ storage nodes store the data nodes' data in original form, anyone of these $k$ storage node can be contacted to collect $1/k$-portion of the original data, for all $\frac{T}{mk}$ groups. This is particularly important where not every sequence of generated data is immediately significant to a data-collector.

Encoding for all groups and all data  nodes yields
$$k\cdot m^2\cdot nk\cdot\frac{T}{mk}\text{ XORs}$$
because i) each node performs encoding, ii) we use a classical matrix multiplication, and iii) all groups are encoded.
From Section~\ref{sect:decoding_complexity} and iterating over all groups, the decoding complexity is $Tmk^2$ XORs in the absence of erroneous storage nodes, and given by
$$Tm(vk^2 +k(n-k)+v^2k+v(n-k)+v^3)\text{ XORs}$$ in the presence of $v$ erroneous storage nodes. Note that IRD is also the only erasure coding scheme to detect and efficiently correct errors. Moreover, IRD adapts decoding computation to the number of erroneous nodes. Finally, neither DEC nor DFC are suitable for real time dependable applications, since they are not deterministic. That is, their ability to decode is probabilistic and cannot be guaranteed.
\section{Implementation and Evaluation}
\label{sect:eval}
We have implemented the proposed and baseline algorithms, where each data node's 
information is a memory buffer in a single machine having
$2.66$ GHz Intel Xeon CPU, $4096$ KB cache and
$2$ GB RAM. A randomly generated message is first
partitioned into either $101$ or $401$ information symbols and then encoded into $n=1023$ coded
symbols of length $10230$ bits. 
Thus, the field size is $2^{10} = 1024$. A
stored symbol is corrupted with an error probability $p$ independently.
Comparing our error-erasure code to either decentralized or fountain erasure codes for error-correction performance is pointless, since these codes
\emph{cannot} feasibly guarantee data availability in the presence of errors. Therefore, in this section, the following three algorithms are considered.
\begin{itemize}
\item{\it BMA} is the Berlekamp-Massey (BMA) algorithm~\cite{MOO05} for
RS decoding. Similar to Algorithm~\ref{algo:retrieval}, BMA {\it progressively}
retrieves data from each storage node and performs decoding until
the decoded symbols passes the CRC checks or failure is declared. However,
decoding cannot be performed incrementally. 
\item{\it BMA-genie} knows \emph{a priori} how many symbols are needed to
successfully decode. BMA-genie decodes only once after retrieving sufficient
number of symbols. Note that BMA-genie is impossible to implement in practice,
and is included for comparison purpose only.
\item{\it IRD} is the proposed progressive data retrieval algorithm. 
\end{itemize}
%
\subsection{Total computation time}
\begin{figure*}[thp]
\begin{center}
\begin{tabular}{cc}
\includegraphics[width=2.5in]{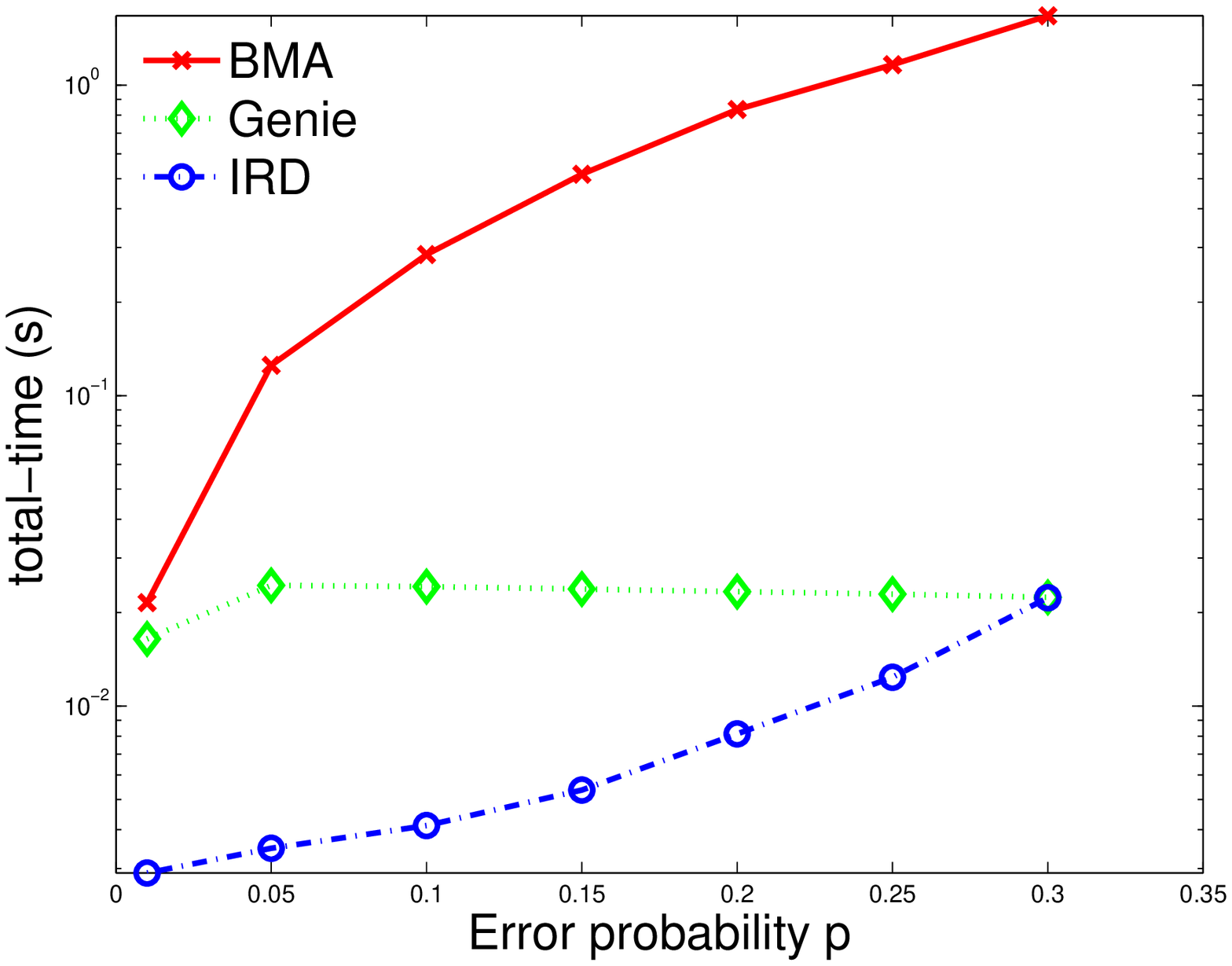} &
\includegraphics[width=2.5in]{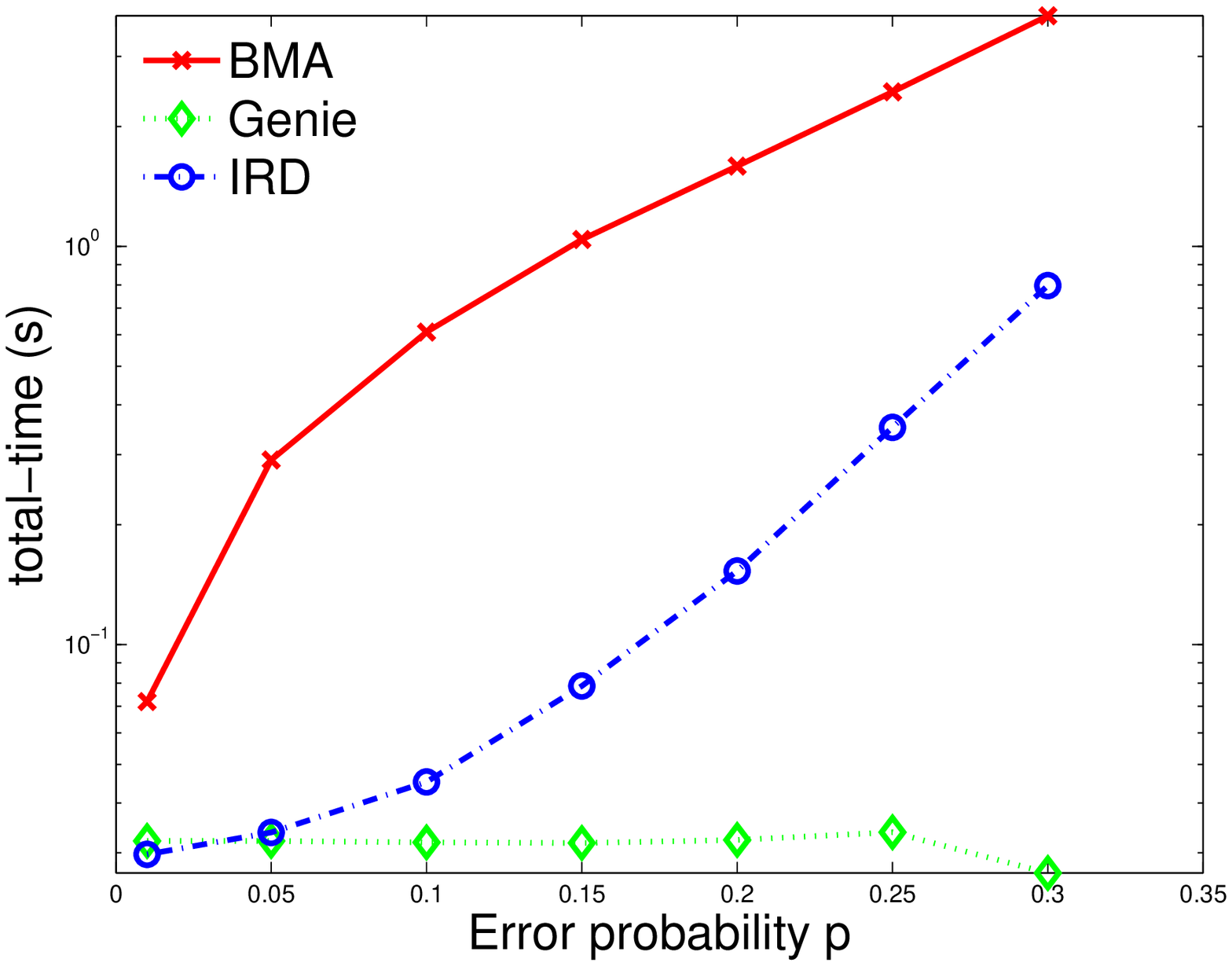} \\
(a) k = 101 & 
(b) k = 401 
\end{tabular}
\caption{Average computation time for $(1023,k)-$MDS decoding against Byzantine node rate, $p$. 
For the given $k$ values, both IRD and BMA cannot successfully decode for $p>0.3$}
\label{fig:alg_time}
\end{center}
\end{figure*}

Figure~\ref{fig:alg_time}(a) and \ref{fig:alg_time}(b) illustrate the computation time (in log scale)
spent in decoding when $k=101$ and $k=401$, respectively. Note that $\hat{k}=k$ in these simulations. The storage overhead
$n/k$ is 10.13 and 2.55 with the maximum number of errors correctable being 461
and 311. From Figure~\ref{fig:alg_time}, we observe that the BMA and IRD
computation time increases  as $p$ increases. But the rate of increment in
IRD is much slower. When $k$ is small or the redundancy is higher
(Figure~\ref{fig:alg_time}(a)), IRD is faster than the genie-aided
BMA. This is because in the genie-aided BMA, the computations of erasure
polynomials (with $O((n-k)^2)$) dominate the decoding time when $p$ is small.
In contrast, IRD does not compute erasure polynomials.

In Section~\ref{sect:complexity}, the data collection costs were shown to depend on 
$\hat{k}$. We quantified the encoding and decoding computational
complexity. 
The evaluation results--for a given Byzantine node rate $p=.2$, $k = 200$ data nodes, and a
$(1023,\hat{k})-$MDS code, where $\hat{k}\in\{50,100,...,550\}$--are
consistent with the analysis in Section~\ref{sect:complexity}: 
encoding computational costs are invariant
of $\hat{k}$, and are relatively insignificant. However, decoding computational costs increase according to Eq.~(\ref{eq:dec_comp}).
%

\subsection{Decoding Breakdown}
We break down the decoding computation time to understand the dominant operations in the algorithms 
as the error probability increases. The break down includes the time 
to find the error-locator polynomial ({\it elp-time}), find the error locations ({\it
chien-time}) and solve for the information polynomial ({\it inv-mat-time}). This
breakdown is also illustrated in
Figure~\ref{fig:rs_decode}, where the $1$st and $2$nd blocks shows the elp-time, while the $3$rd
and $4$th blocks give the chien-time and inv-mat-time, respectively. 

\begin{figure*}[thp]
\begin{center}
\begin{tabular}{ccc}
\includegraphics[width=2.3in]{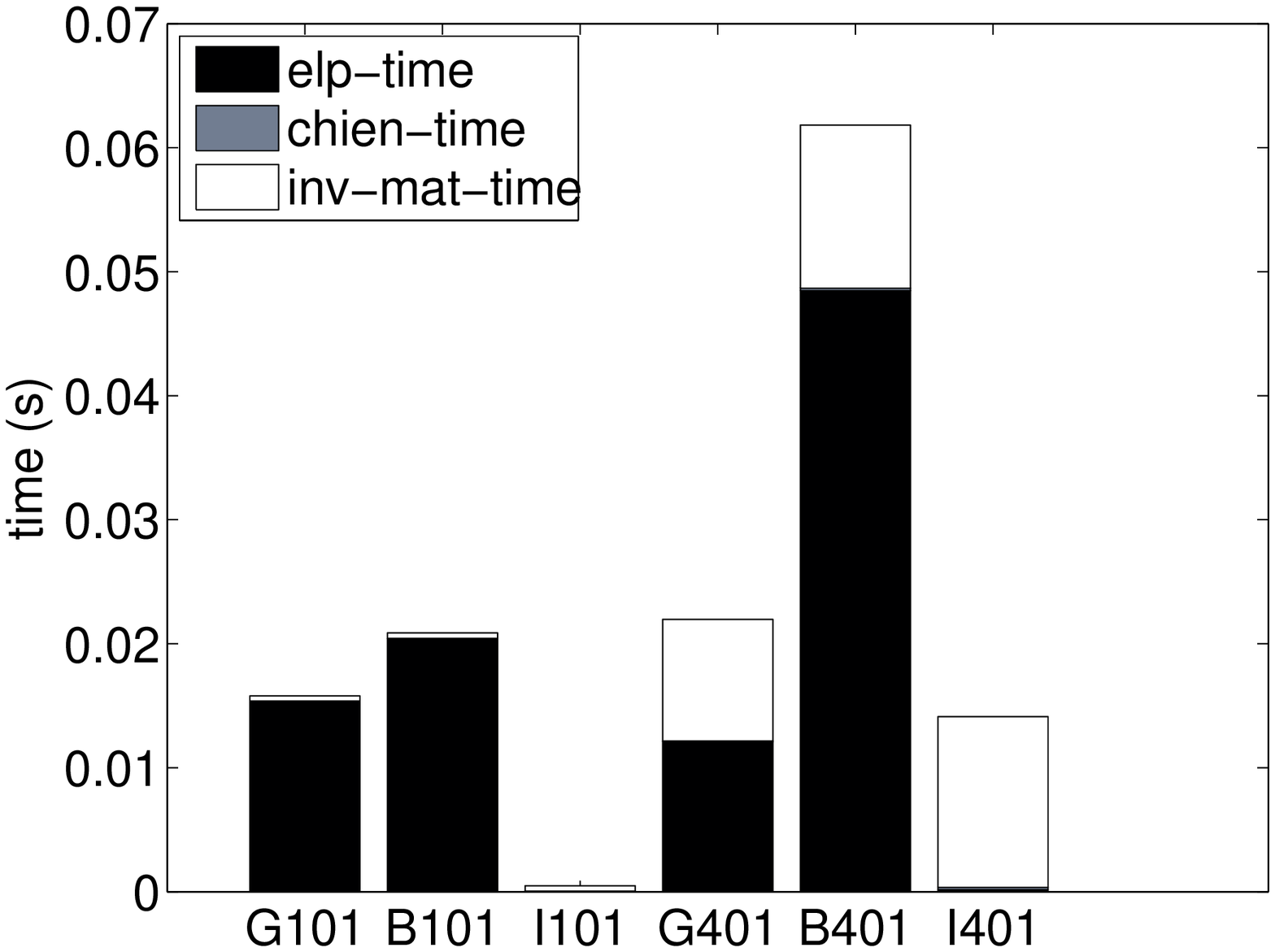} &
\includegraphics[width=2.3in]{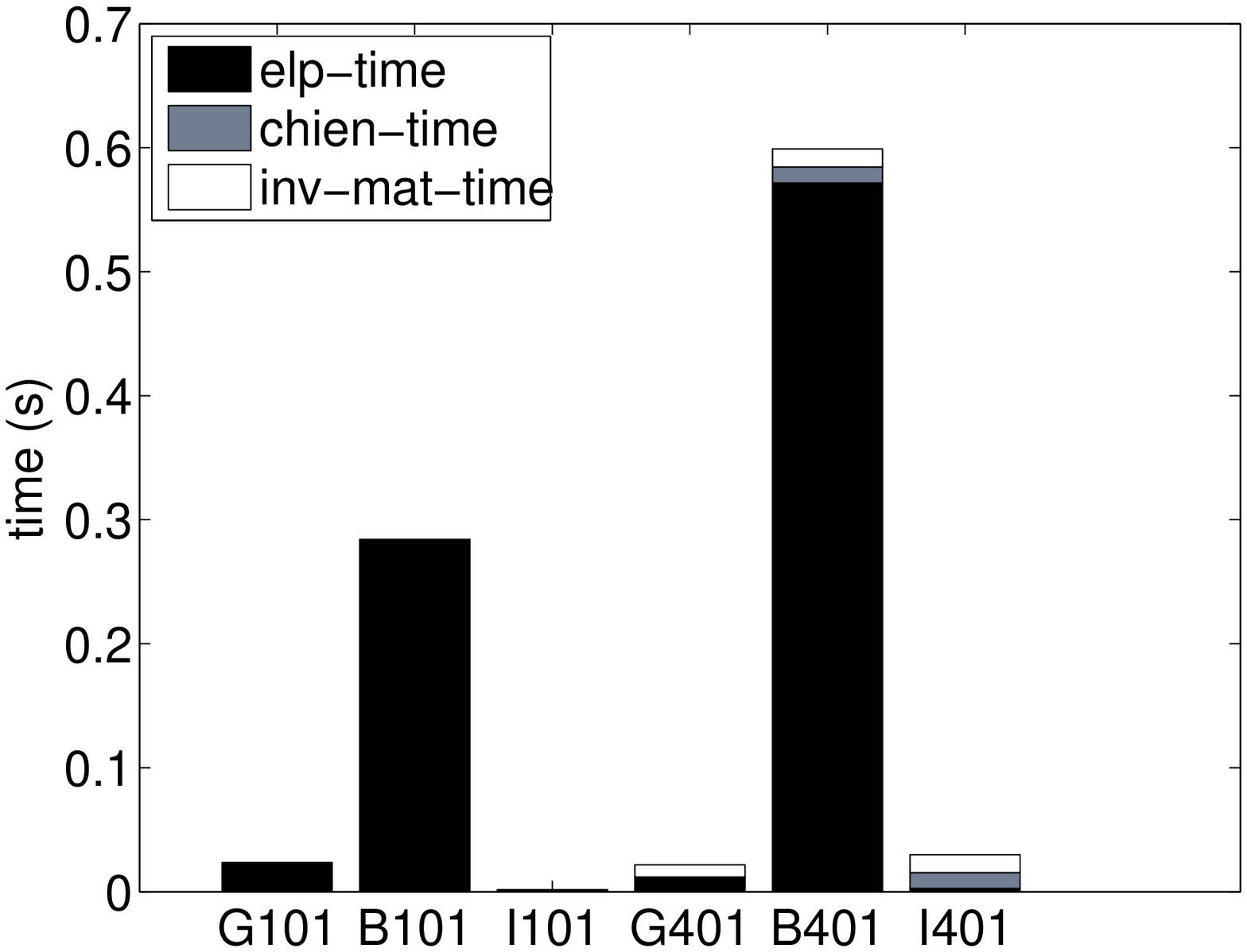} &
\includegraphics[width=2.3in]{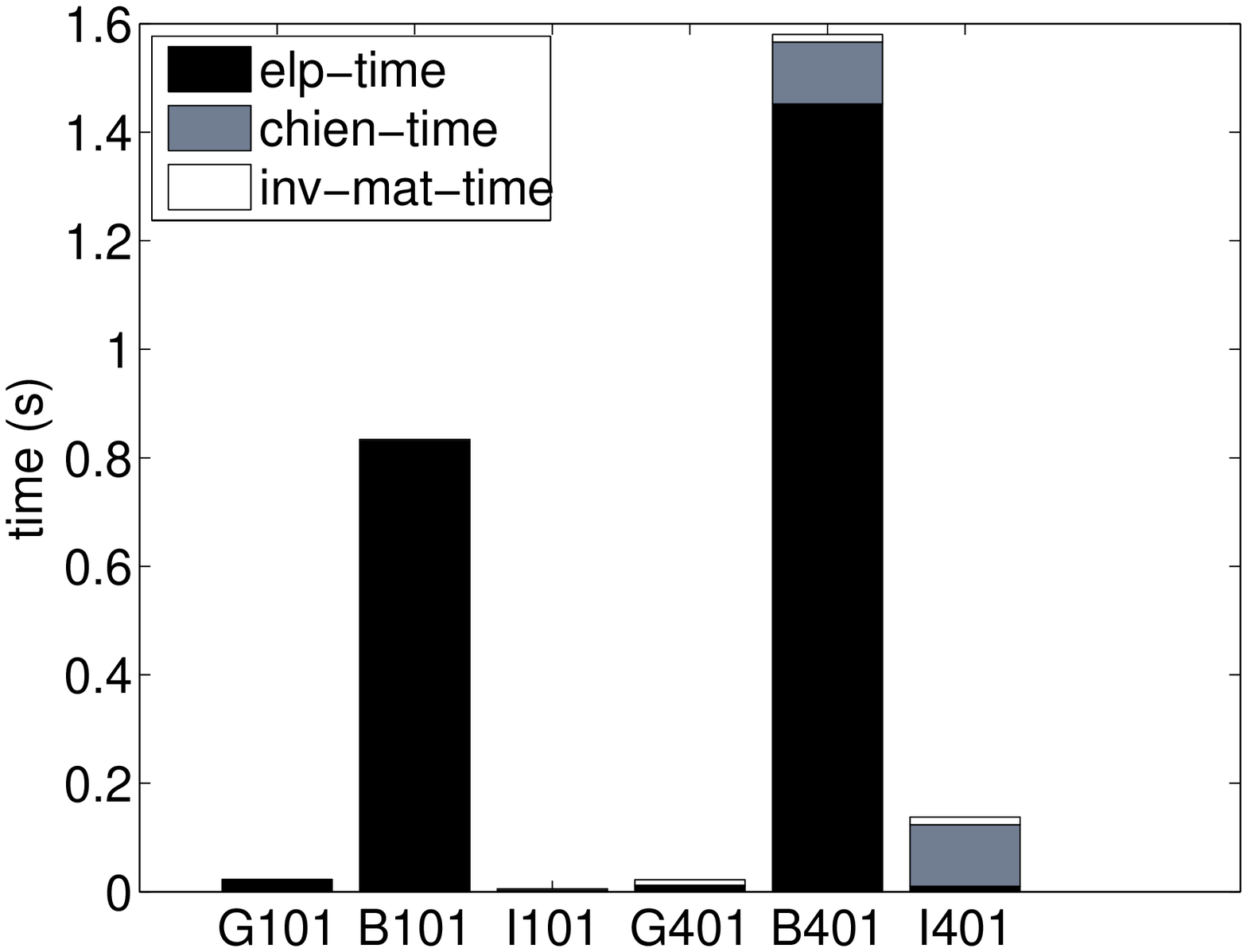} \\
(a) $p = 1\%$ &
(b) $p =10\%$ &
(c) $p = 20\%$ 
\end{tabular}
\caption{Average computational time breakdown for decoding one $(1023,k)-$MDS codeword,
$k\in\{101,401\}$. Because IRD progressively decodes, its performance
does not deteriorate with an increasing Byzantine node rate, $p$.}
\label{fig:breakdown}
\end{center}
\end{figure*}

When the error probability is low (Figure~\ref{fig:breakdown}(a)), computation
of error-locator polynomials dominates for small $k$, while the
matrix inversion time becomes significant when $k$ is large. In our
implementation, the cost of a matrix inversion is quadratic in the number of
symbols decoded. Chien search though asymptotically is the most time consuming
procedure, it can be performed quite fast. When the error
probability is high (Figure~\ref{fig:breakdown}(b),(c)), computation of error location polynomials
dominates except in IRD. Also, from Figure~\ref{fig:breakdown},
we observe that the computation time in matrix inversion is almost
negligible (on the order of tens of milliseconds) in BMA and IRD, and
is comparable to that in BMA-genie (recall that BMA-genie knows the number of
errors in advance and thus performs matrix inversion only once). This is
because even though there are more errors with larger $p$ (and thus more
iterations), the decoding algorithm is likely to fail in or before Chien search
(e.g., Algorithm~\ref{algo:incremental} (Line~\ref{chien})). Thus, in most
cases, BMA and IRD perform matrix inversion once.

From the experiments, IRD is more efficient because it utilizes intermediate results from
previous iterations. Up to $20$ times speed up can be attained, relative to
BMA.
%
%
\section{Conclusions}
\label{sect:conc}
\vspace{-0.25em}
We have developed a communication-optimal algorithm to guarantee data dependability 
and availability for distributed storage systems, in the midst of Byzantine-faulty and crash-stop nodes.
The communication cost for data retrieval is minimized by utilizing intermediate computation 
results and collecting only the minimum data required 
for successful data reconstruction. The efficient encode and decode
primitives serve as a fundamental building block for distributed dependable
storage systems. An analytical model to evaluate the communication complexity of our incremental data retrieval algorithm is provided.
Moreover, our previous work restricted 
$\hat{k}=k$, a constraint that is unrealistic for distributed networked storage systems. In this paper, the constraint is eliminated, and 
$\hat{k}$ is invariant of the number of data nodes, $k$. Finally, our implementation
results show that our progressive scheme outperforms the state-of-the-art scheme by a factor of $20$ in computation costs.
Moreover, they are consistent with our analytical model, for any $\hat{k}$ and any
Byzantine node rate.
%
\appendix[Proof of Theorem~\ref{thm:average_access}]
\label{app:proof}
Let $A_v$ be the event that there are $v$ compromised
storage nodes in the network, and $B_i$ be the event that the error-erasure
decoding algorithm  executes $i$ times when it completes {\it successfully}.
Note that $i$ also indicates how many errors the error-erasure decoding
algorithm has corrected since our proposed scheme asks for extra data to
correct one more error in each iteration.  

Therefore, the average number of accesses
is given as,
\begin{eqnarray}
\nonumber
& & \bar N(n,\hat{k})
\\ 
\nonumber
&=&\sum_{v=0}^{n-\hat{k}}\Pr(A_v)\sum_{i=0}^{\min(v, \lfloor{\frac{n-\hat{k}}{2}}\rfloor, n-v-\hat{k})}(\hat{k}+2i)\Pr(B_i|A_v) \\
\nonumber
&+& \sum_{v=0}^{n-\hat{k}}n\Pr(A_v)\left(1-\sum_{i=0}^{\min(v, \lfloor{\frac{n-\hat{k}}{2}}\rfloor, n-v-\hat{k})}\Pr(B_i|A_v)\right) \\
&+ &\sum_{v=n-\hat{k}+1}^{n}n\Pr(A_v)~.
\label{average}
\end{eqnarray}
The first term gives the average number of accesses  in a  successful run. The third and second  terms
correspond to the first and second failure cases discussed in Section~\ref{sect:analysis}, respectively.  $\Pr(A_v)$
is simply given as $${n\choose v}p^v(1-p)^{n-v}~.$$

To determine $\Pr(B_i|A_v)$, we first derive $\Pr(B_0|A_v)$, i.e., the
probability that only erasure decoding is needed. Clearly, $B_0$ occurs if
the first $\hat{k}$ copies of data are from healthy nodes. Therefore, $\Pr(B_0|A_v) =
{n-v \choose \hat{k}}/{n \choose \hat{k}}.$ Note this results holds even when $v >
\lfloor{\frac{n-\hat{k}}{2}}\rfloor$.
			
When $i > 0$, let
$A(l)$ and $B(l)$ represent the number of erroneous data and correct data
received at the data collector after accessing $l = \hat{k}+2i$ live storage nodes. Clearly,
$A(l) + B(l) = l$.  The event $B_i$ occurs under the following conditions, (i)
$A(2i+\hat{k}-1) = i$ and $B(2i+\hat{k}-1) = i + \hat{k} - 1$; {\it and} (ii) $B(l) - A(l) <
\hat{k}$,$\forall l \le 2i+\hat{k}-1$; {\it and} (iii) $A(2i+\hat{k}) =  i$ and $B(2i+\hat{k}) = i+\hat{k}$.
Evolution of $A(l)$ and $B(l)$ can be modeled as a lattice path from the origin
$(0,0)$ to $(i, i+\hat{k})$ in the A-B coordinate system, recording the running
totals as more nodes are accessed. Condition (i) implies the
path has to go through the point $(i, i+\hat{k}-1)$.  Condition (ii) implies that the path {\it
never} intersects the line $y = x + \hat{k}$ except in the $i$th step.  Condition (iii) implies the last data
retrieve needs to be from a healthy node. The lattice path is the result of
directional random walks, where each move is conditionally independent of the
previous moves given the current coordinates. At step $l$, the probability $A(l+1) = A(l) + 1$ and $B(l+1) =
B(l)$ is given by
$$\frac{v - A(l)}{n - l}~,$$
since there are $n-l$ and
$v-A(l)$ remaining nodes and compromised nodes, respectively. The probability
$A(l+1) = A(l)$ and $B(l+1) = B(l)+1$ is given by
$$\frac{n - v - B(l)}{n - l}$$
since there are $n-l$ and $n-v-B(l)$ remaining nodes and healthy nodes,
respectively. Therefore, the probability for Condition (iii) to hold on Condition  (i) is
$$\frac{n - v - (i+\hat{k}-1)}{n - (2i+\hat{k}-1)}~.$$ The probability for Condition (i) to hold is simply,
$$\frac{{n-v\choose i+\hat{k}-1}{v \choose i}}{{n \choose 2i+\hat{k}-1}}~.$$

Now, what remains to be derived is the probability of (ii) on condition (i).
We use the bijection arguments due to Antone Desir\'{e} Andr\'{e}~\cite{west:01}, and count
instead the number of ``bad" paths that cross the line $y = x+\hat{k}$. Consider the point $(i-1,i+\hat{k})$ which is the reflection of the point $(i,i+\hat{k}-1)$ along the line $y=x+\hat{k}$. Clearly, the point $(i - 1, i+\hat{k})$ is above the line $y = x +\hat{k}$. Thus, all paths from $(0,0)$ to $(i - 1, i+\hat{k})$ must hit the line $y = x + \hat{k}$ at least once. Consider the
first time such a path $\PP$ hits the line $y = x + \hat{k}$ at $(j, j+\hat{k})$. After this point, the remaining number of correct data  along this path is $i-1-j$ and that of erroneous data is $i-j$. Now consider a path $\PP'$ coinciding with $\PP$ up to point $(j,j+\hat{k})$ and afterward it has exact opposite branches to $\PP$. That is,
the results of data retrieval are switched afterward, namely, all data from compromised data
are counted toward $B(l)$ and all data from healthy nodes are counted toward
$A(l)$. As a result, $A(l) = j + i-j = i$ and $B(l) = j + \hat{k}
+ i -1- j = i+ \hat{k}-1$ for $\PP'$.  Thus, for any path reaching $(i - 1, i+\hat{k})$, reflecting the remainder of the path
after it first hits $y = x + \hat{k}$ yields a ``bad" path to $(i, i+\hat{k}-1)$. Similarly, every ``bad" path to  $(i, i+\hat{k}-1)$ has a corresponding such path to $(i - 1, i+\hat{k})$ that intersects with the line $y = x + \hat{k}$ by construction. This
establishes a bijection between the set of ``bad" paths to $(i, i+\hat{k}-1)$ and
the set of paths to $(i - 1, i+\hat{k})$. Clearly, there are ${2i+\hat{k}-1\choose i-1}$
``bad" paths. Therefore, the probability for Condition (ii) to hold on Condition (i) is,
\begin{equation}
\frac{{2i+\hat{k}-1\choose i} - {2i+\hat{k}-1\choose i-1}}{{2i+\hat{k}-1\choose i}} = \frac{\hat{k}}{i+\hat{k}}~.
\end{equation}

To this end, we obtain the probability that the error-erasure decoding
algorithm stops at the $i$th iteration when there are $v$ compromised nodes as
follows, $\forall i > 0$,
\begin{equation}
\Pr(B_i|A_v) =  \frac{{n-v\choose i+\hat{k}-1}{v \choose i}}{{n \choose 2i+\hat{k}-1}}\times \frac{\hat{k}}{i+\hat{k}} \times \frac{n - v - (i+\hat{k}-1)}{n - (2i+\hat{k}-1)}~.
\end{equation}

We now have the average number of accesses and the probability of
successful decoding given in Eq.~\eqref{eq:average} and
Eq.~\eqref{eq:sucprob}, respectively.
%
%
\section*{Acknowledgment}
Han's work was supported  by 
the National Science Council of Taiwan, under grants NSC
96-2221-E-305-002-MY3 and his
visit to LIVE lab at University of Texas at Austin. Omiwade and Zheng's work is supported in part by NSF CNS 0546391.
\bibliographystyle{IEEE}
\bibliography{network,nfs}
\end{document}